\documentstyle[eqsecnum,aps,epsf]{revtex}
\begin{document}

\title{The imposition of Cauchy data to the Teukolsky equation II:\\
Numerical comparison with the Zerilli-Moncrief approach to black hole
perturbations}
\author{Manuela Campanelli\thanks{
Electronic address: manuela@mail.physics.utah.edu}}
\address{Department of Physics, University of Utah, 201 JBF, 
         Salt Lake City, UT 84112, USA}
\author{William Krivan\thanks{
Electronic address: krivan@mail.physics.utah.edu}}
\address{Department of Physics, University of Utah, 201 JBF, 
         Salt Lake City, UT 84112, USA\\
    and Institut f\"ur Astronomie und Astrophysik, Universit\"at 
        T\"ubingen, D-72076 T\"ubingen, Germany}
\author{Carlos O. Lousto\thanks{
 Electronic address: lousto@iafe.uba.ar}}
\address{Instituto de Astronom\'\i a y F\'\i sica del Espacio,
         Casilla de Correo 67, Sucursal 28, (1428) 
         Buenos Aires, Argentina}

\date{\today}
\maketitle

\begin{abstract}
We revisit the question of the imposition of initial data representing
astrophysical gravitational perturbations of
black holes. We study their dynamics for the case of nonrotating
black holes by numerically evolving the Teukolsky equation in the time
domain.
In order to express the Teukolsky function $\Psi$ explicitly in terms of
hypersurface quantities, we relate it to the Moncrief waveform
$\phi_{\text{M}}$ through a Chandrasekhar transformation in the
case of a nonrotating black hole.
This relation between $\Psi$ and $\phi_{\text{M}}$ holds for any constant
time hypersurface and allows us to compare the computation of the
evolution of Schwarzschild perturbations by the Teukolsky and by the Zerilli
and Regge-Wheeler equations. We explicitly perform this comparison for
the Misner initial data in the close limit approach. We evolve numerically 
both, the Teukolsky (with the recent code of Ref.\ \cite{KLPA97})
and the Zerilli equations, finding complete agreement in resulting waveforms
within numerical error.
The consistency of these results further supports the correctness of the
numerical code for evolving the Teukolsky equation as well as 
the analytic expressions for $\Psi$ in terms only of the three-metric and
the extrinsic curvature.

\end{abstract}
\pacs{04.30.Db; 04.70.Bw}

\section{Introduction and overview}

Binary black holes provide, in principle, one of the 
strongest sources of the gravitational radiation to be
observed by the detectors currently under construction.
Black holes have proved to be very elusive to standard astronomical methods
of detection. This is a consequence of the fact that
astrophysical black holes, being one of the simplest objects in
nature, interact only gravitationally with the rest of the universe.
Thus, gravitational wave observatories are particularly suitable to 
detect them.
The confirmation of black hole existence in nature
can be considered to be as great a discovery as 
the confirmation of the existence of gravitational radiation.

This exciting new experimental-observational situation has taken theory a bit
by surprise. In the early seventies a great deal of effort has been devoted
to study gravitational perturbations by means of the Post-Newtonian
and perturbative approaches. In 1979 Smarr\cite{S79}
presented his pioneer work on
the numerical attack to the full Einstein equations. Subsequently 
the field developed rather slowly, 
in part due to a shift of the research interests
toward other fields and in part by the lack of concrete experimental support.
The situation changed dramatically in the early nineties with the launch of
the interferometric detectors more or less simultaneously in Europe and the
USA. Renewed efforts have been devoted to the full numerical approach in the
form of the ``Grand Challenge alliance''\cite{GrandChallenge},
to the post-Newtonian approximation\cite{DP97},
and the perturbative approach. 
It was the perturbative method which in 1994 lead to the remarkable ``close 
limit approximation''\cite{PP94}.
Its, perhaps unexpected, success in describing the final stage of the
binary black hole coalescence in terms of a single perturbed black hole was
somewhat perturbing\cite{GNPP96}. A complementary approach to this
approximation is to consider not two black holes of similar masses (here the
close limit gives its optimal approximation), but instead one much
less massive than the other. This, again, allows us to treat the problem as 
a perturbation around a single black hole, although here the perturbation
parameter is no longer the separation of the holes, but the mass ratio.
This problem was recently revisited and solved for a finite initial separation
of the holes\cite{LP97a,LP97b,LP98}.
However, both the close and the particle limits have been studied
as perturbations around nonrotating (Schwarzschild) black holes. We know that
astrophysical black holes are most likely to be rotating, possibly with
considerable angular momentum.
This means that we should take as the
background metric the Kerr rather than the Schwarzschild geometry.

Using the Newman-Penrose formalism, Teukolsky\cite{T73}
succeeded in giving a decoupled
equation for perturbations around a Kerr background. Notably, this
equation can be separated in all its variables. The Teukolsky equation has
been studied and solved in the {\it frequency} domain for the gravitational
radiation generated by a particle
infalling from infinity or in circular orbit around a Kerr hole\cite{D78}. 
Nakamura and Sasaki\cite{SN82}
have given a modified version of the radial Teukolsky
equation in the frequency domain that is better behaved
(see Ref.\ \cite{CL97} though) and suitable for
numerical integration. Different trajectories of a particle infalling
towards a Kerr black hole have been studied in this way\cite{ponjas}.
The disadvantage of this approach, however, is that, in general,
it is much more efficient to perform the numerical integrations in the
{\it time} domain. This was concretely our experience in Ref.\ \cite{LP97b},
where in order to reproduce in the time domain the results of Ref.\ 
\cite{LP97a}
obtained in the frequency domain, the running time of the program
was reduced by three orders of magnitude.

A successful code for the numerical integration of the Teukolsky equation
in the time domain was built up only very recently \cite{KLPA97}.
It has been used to study quasinormal mode excitation, superradiance, and 
the power law behavior of the (late time) gravitational
radiation tails on the Kerr background. Also, consistency checks
similar to some of the material presented here, were performed during the 
development of the code.
Given the success of the close limit
approximation for the head-on collision of black holes, it is very
interesting
to study the same problem in the astrophysically more realistic situation of
two merging black holes in an inspiral orbit. Presumably, this situation
can be studied in terms of a single, highly rotating, perturbed Kerr black
hole. To complete the solution of this problem, we have to provide the 
evolution code with consistent initial data. There are two kind of problems
that we have to address now. 
Firstly, {\it what} initial data do we take? Here the problem is that, so far,
all initial data available in the literature representing
binary black holes assumed the initial three-geometry to be
conformally flat. But since a $t$=constant slice of the Kerr geometry
(at least in 
Boyer-Lindquist coordinates) is not conformally flat, we cannot
consistently use those initial data. Only very recently a first step towards
a different approach has been given by assuming the perturbed
metric to have the Kerr-Schild form \cite{BIMW98}. A second practical problem
appears even if one had consistent initial data. {\it How} to impose
them in terms of
the initial Teukolsky waveform, $\Psi$, in order to start the evolution?
We have begun to discuss this problem in Ref.\ \cite{CL98}. The problem here
is that $\Psi$ is built up as a contraction
of the perturbed Weyl tensor with
the background (Kerr) null tetrad. This contraction, when written in terms of
metric perturbations,
depends explicitly not only on the hypersurface quantities
(three-metric ${}^{(3)}g_{ij}$ and extrinsic curvature $K_{ij}$), 
but also on the lapse function $N$ and the shift vector
$N^i$. We then have to rewrite $\Psi$ in
terms of ${}^{(3)}g_{ij}$ and $K_{ij}$ for
$t=0$ (i.e.\ on the initial hypersurface) 
by making use of not only the constraints but
also the evolution equations. In practice this task proved to be nontrivial.
This is in contrast with the description, mostly after the
work of Zerilli\cite{Z70} and of Moncrief\cite{M74}, of metric perturbations
around a Schwarzschild hole. In this case,
the Moncrief waveform $\phi_{\text{M}}$ depends
explicitly on the perturbed three-geometry only
and $\partial_t\phi_{\text{M}}$ can be written 
in terms of the extrinsic curvature only. (There are actually two waveforms
representing even and odd parity perturbations and they correspond, 
respectively, to the real and imaginary parts of the Teukolsky 
waveform $\Psi$.) Our aim is
to bring the Newman-Penrose-Teukolsky formulation to an equally nice footing.
One first step that can be done is, at least in the limit of a
nonrotating background, to
find a relation between $\Psi$ and $\phi_{\text{M}}$
on the initial hypersurface. Such
a relation was found in Ref.\ \cite{CL98} making use of the simplifications
introduced by choosing the Regge-Wheeler gauge and the possibility to restore
the gauge invariance at the end of the calculation.

When one speaks of transformations between solutions of Teukolsky and
Zerilli (or Regge-Wheeler) equations, one immediately associates with them
the name of Chandrasekhar for his contributions to the understanding of
these relations. In the next section, we extend Chandrasekhar's 
transformations to the time domain in order to relate
$\Psi$ to $\phi_{\text{M}}$, and then be able
to rewrite the Teukolsky waveform exclusively in terms of 
hypersurface quantities.
We test this relation, that holds for every hypersurface, by numerically
evolving initial data corresponding to the close limit in two independent
ways: by the Teukolsky equation with the code detailed in Ref.\ \cite{KLPA97}
and by the Zerilli equation. We then compare waveforms, making use of the above
relations, at later times, and check their agreement numerically.
In turn, this provides further support about the correctness of the numerical
code for evolving the Teukolsky equation. The numerical results are displayed
in Sec.\ III. We conclude this paper with a brief
discussion of our results and their generalization
to the rotating background case.

\section{Chandrasekhar transformation in the time domain}

\subsection{Teukolsky equation}

Let us briefly review the Newman-Penrose-Teukolsky description of perturbations
around the Kerr metric. Gravitational perturbations with spin-weight 
$s=\pm2$ are compactly written in terms of contractions of the Weyl tensor
\begin{equation}  \label{psi}
\Psi (t,r,\theta ,\varphi )=\left\{ 
\begin{array}{ll}
\rho^{-4}\Psi_4\equiv -\rho^{-4}C_{n\bar mn\bar m} & {\rm for}~~s=-2
\\ 
\Psi_0\equiv -C_{lmlm} & {\rm for}~~s=+2~
\end{array}
\right. ,  \label{RH}
\end{equation}
where an overbar means complex conjugation, $\rho \equiv 
1/(r-ia\cos \theta )$,
and we have considered Boyer-Lindquist $(t,r,\theta,\phi)$
coordinates. This field
represents either the outgoing radiative part of the perturbed Weyl tensor, 
($s=-2$), or the ingoing radiative part, ($s=+2$).
The components of the Kinnersley null tetrad \cite{K69} are given by
\begin{mathletters}
\begin{eqnarray}
   (l^{\alpha}) &=& \left( \frac{r^2+a^2}{\Delta},1,0,\frac{a}{\Delta} 
                  \right) \; , \\
   (n^{\alpha}) &=& \frac{1}{2\;\!(r^2+a^2\cos^2\theta)} \,
                  \left( r^2+a^2,-\Delta,0,a \right) \; , \\
   (m^{alpha}) &=& \frac{1}{\sqrt{2}(r+ia\cos\theta)}\,
                  \left( ia\sin\theta,0,1,i/\sin\theta \right)\; . 
\end{eqnarray}
\end{mathletters}

The Weyl scalars then satisfy the Teukolsky equation
\begin{eqnarray}
&&\ \ \Biggr\{\left[ a^2\sin^2\theta -\frac{(r^2+a^2)^2}\Delta \right]
\partial_{tt}-\frac{4Mar}\Delta \partial_{t\varphi }-2s\left[ (r+ia\cos
\theta )-\frac{M(r^2-a^2)}\Delta \right] \partial_t  \nonumber \\
&&\ \ +\,\Delta^{-s}\partial_r\left( \Delta^{s+1}\partial_r\right) +
\frac 1{\sin \theta }\partial_\theta \left( \sin \theta \partial_\theta
\right) +\left( \frac 1{\sin^2\theta }-\frac{a^2}\Delta \right) 
\partial_{\varphi \varphi }  \label{master} \\
\ &&+\,2s\left[ \frac{a(r-M)}\Delta +\frac{i\cos \theta }{\sin^2\theta }
\right] \partial_\varphi -\left( s^2\cot^2\theta -s\right) \Biggr\}\Psi
=4\pi \Sigma T \;,  \nonumber
\end{eqnarray}
where $M$ is the mass of the black hole, $a$ its angular momentum per unit
mass, $\Sigma  \equiv r^2+a^2\cos^2\theta$, and $\Delta 
\equiv  r^2-2Mr+a^2$.
The source term $T$ is built up from the energy-momentum tensor\cite{T73}.

\subsection{Zerilli equation and Moncrief waveform}

There is a historically independent way of computing gravitational
perturbations around a Schwarzschild black hole (it can be likewise extended
to any spherically symmetric background), developed mainly by
Regge and Wheeler\cite{RW57}, Zerilli\cite{Z70}
and Moncrief\cite{M74}. In that formalism, 
the two degrees of freedom of the graviton are
represented by two scalar quantities
$\phi_{\text{M}}$, the even and odd parity
waveforms. After decomposition of the angular part in terms of spherical
harmonics $Y_{\ell}^m(\theta,\phi)$, $\phi_{\text{M}}$ satisfies 
a wave equation
\begin{equation}\label{rtzerilli}
-\frac{\partial^2\phi_{\text{M}}}{\partial t^2}
+\frac{\partial^2\phi_{\text{M}}}{\partial r*^2}-V_{\ell}(r)\phi_{\text{M}}=
{\cal S}_\ell (r,t) \; .
\end{equation}
Here $r^*\equiv r+2M\ln(r/2M-1)$, ${\cal S}_\ell$ is the contribution of the
source terms, and $V_\ell$ is the potential due to the curved background
(slightly different for the even and odd parity waves).

The even and odd parity waveforms in terms of metric perturbations in the
Regge-Wheeler notation\cite{RW57} take the form

\begin{equation}
  \phi_{\text{M}}(r,t)=\left\{
\begin{array}{ll}
\frac{r}{{\lambda}+1}\left[
    K+\frac{r-2M}{\lambda r+3M}\left\{ H_2-r\partial K/\partial r
    \right\} \right]
+\frac{(r-2M)}{\lambda r+3M}\left(r^2\partial G/\partial r-2h_1
\right)\\ \\
\frac {1}{r}\left(1-\frac{2M}r\right) \left[ h_1+\frac 12\left(
\partial_r h_2-\frac 2rh_2\right) \right]
\end{array}\right.,
\label{psidef}
\end{equation}
where \begin{equation}
{\lambda} \equiv (\ell+2)(\ell-1)/2\ \;,
\end{equation}
and  $\ell$ is the multipole index.
The field $\phi_{\text{M}}$ explicitly depends only on the three-geometry.
Likewise, one can write
$\partial_t\phi_{\text{M}}$ exclusively in terms 
of the extrinsic curvature, using the same functional 
form of $\phi_{\text{M}}$ as above \cite{AP96a}:
\begin{equation}\label{formpsidot}
\partial_t\phi_{\text{M}}=-2\phi_{\text{M}}\left\{(1-2M/r)^{1/2}
\delta K_{ij}, \partial_r\left[(1-2M/r)^{1/2}\delta K_{ij}
\right]\right\}\ \;.
\end{equation}

\subsection{Relation between waveforms}

The Teukolsky function can be decomposed into angular modes\cite{CL98}
\begin{equation}
\Psi_4(t,r,\theta,\phi)=\sum_{\ell m}\Psi_4^\ell(t,r)\ 
{}_{-2}Y_\ell^m(\theta,\phi), 
\end{equation}
where we have made use of the fact that the spin-weighted spheroidal harmonics,
in the case $a\omega=0$, reduce to the spin-weighted spherical harmonics,
i.e. $e^{im\phi}{}_{-2}S_\ell^m(\theta,a\omega=0)=
{}_{-2}Y_\ell^m(\theta,\phi)$.

Chandrasekhar transformations provide a differential operator
(first order in frequency domain,
but second order in time domain) that links
solutions of the Teukolsky equation with
solutions of the Zerilli or Regge-Wheeler equations and vice versa.
Chandrasekhar found his transformations in the
frequency domain, but in the nonrotating background case  they can be easily
extended to the time domain as (see Eqs.\ (3.353) and (3.345) of
Ref.\ \cite{C83})
\begin{equation}\label{chandra}\left\{
\begin{array}{rr}
\text{Re }{\Psi}_4(t,r,\theta,\phi)\\
-\text{Im  }{\partial_t\Psi}_4(t,r,\theta,\phi)
\end{array}\right\}
=\sum_{\ell m}\frac{C}{r} \left[ -2 \partial_{tr^*} 
+ 2 \partial_{r^*r^*}-V^\pm(r)
+W^\pm(r)\left( \partial_{r^*}
\mp\partial_t\right)\right] \,\phi_{\text{M}}
\, {_{-2}Y_\ell^m}(\theta,\phi) \; ,
\end{equation}
where $C=-\sqrt{(\ell-1)\ell (\ell+1)(\ell+2)} /16 $.
This equation has to be understood as the real part of $\Psi_4$
being equal to the even parity counterpart (labeled as ``+''), and
the imaginary part of $-\partial_t\Psi_4$ equal to the odd parity terms
(labeled as ``$-$''). In the above equation $V^\pm(r)$ denote the Zerilli 
\cite{Z70} and Regge-Wheeler \cite{RW57} potentials, respectively, and
\begin{equation}
W^+(r)=2\,\frac{\lambda r^2 - 3\lambda M r-3 M^2}
                    {r^2 \left(\lambda r+3M\right)}\ \ ,\ \ \ 
W^-(r)=\frac{2(r-3M)}{r^2} \; .
\end{equation}
Chandrasekhar also gives the inverse transformation for $\phi_{\text{M}}$
in terms of a differential operator acting on $\Psi_4$ (see Eqs.\ (3.319) of
Ref.\ \cite{C83}).

We have checked analytically, in the close limit case (see below), that
Eq.\ (\ref{chandra}) leads to exactly the same relation as Eq.\ (3.6) of
Ref.\ \cite{CL98} for all times.

\section{Numerical Tests}

\subsection{Close limit Approximation}

One of the more outstanding results of perturbation theory in the last years
has been its application to the so called ``close limit approximation''. The
basic idea is to regard 
the collision of two black holes in its merger
stage as a {\it single} perturbed black hole. When applied to two equal mass
holes, this approximation gave excellent agreement with the full,
nonperturbative, numerical computations, even going to not so small
separations\cite{PP94}. The simplest application of this method considers
two black holes in head-on collision. The resulting final black hole
will possess no spin and can then be studied as a perturbation around a
Schwarzschild metric. For the two black holes starting from rest at
given separation, we have analytic expressions for the conformally flat initial
geometry. In the Regge-Wheeler notation for the metric
perturbations
the only nonvanishing components of the three-metric, $H_2=K$, are
given by
\begin{equation}
H_2(t=0,r,\theta)=\frac{2M/R}{\left(1+\frac{M}{2R}\right)}
\sum_{\ell=2,4,...}^{\infty}\sqrt{\frac{4\pi}{2\ell+1}}
(Z_0/R)^\ell Y_\ell^0(\theta)\; .
\end{equation}
where $Z_0$ is half the distance between the two (equal masses) holes in
the conformal space. The above equation holds for the Brill-Lindquist
data\cite{BL63}.
The corresponding analysis for the Misner is equivalent to the one above
replacing $(Z_0/M)^\ell\to4\kappa_\ell$.
The results of the Brill-Lindquist case\cite{BL63}
can be compared with those of Ref.\ \cite{AP96b}, while for the case of
Misner data\cite{M60} one can see Refs.\ \cite{PP94,APPSS95}.
For the case of two different masses, see Ref.\ \cite{NP97}.

To compute $\phi_{\text{M}}$ from Eq.\ (\ref{psidef}) we further make the
identification of the conformal radial coordinate with the
Schwarzschild isotropic coordinate, i.e.
\begin{equation}\label{Rdef}
R \equiv \frac{1}{4} \left( \sqrt{r} + \sqrt{r-2M} \right)^2 \; .
\end{equation}

This allows us to compute the form of
$\phi_{\text{M}}$ (here only even parity waves are
generated) and from $\phi_{\text{M}}$  we compute
$\Psi_4$ making use of the relations
(\ref{chandra}). Also, in this case
\begin{equation}
\label{iniPhidotexplicit}
\partial_t \Psi_4 (t=0) =  -\frac{2M}{r^2} \Psi_4 (t=0)\; .
\end{equation}
We now have the explicit form of the initial data for evolving the Teukolsky
equation.

\subsection{Numerical results}

While gravitational perturbations can be described in terms of either
sign of the spin-weight parameter $s$ with $|s|=2$,
it was found more convenient from the numerical point of
view to deal with the $s=-2$ waveform (Refs.\ \cite{KLPA97,PT73}). In
particular, we make a further rescaling of $\Psi_4$ and mode decomposition
in the azimuthal coordinate by defining
\begin{equation}
\Phi(t,r^*,\theta) \equiv \frac{e^{-i m
\tilde\phi}}{\rho^4r^3}\,\Psi_4(t,r^*,\theta,\phi)\;,
\end{equation}
where
\begin{equation}
d\tilde\phi \equiv d\phi+\frac{a}{\Delta}\, dr \; , 
\end{equation}
\begin{equation}
\frac{dr^*}{dr} \equiv \frac{r^2+a^2}{\Delta} \; .
\end{equation}
The outgoing part of $\Phi$ satisfies the asymptotic conditions
\begin{eqnarray}\label{phiasymp}
\lim_{r^* \to +\infty}{|\Phi|} & \sim & 1 \; , \\
\lim_{r^* \to -\infty}{|\Phi|} & \sim & 1 \; ,
\label{asymp-out}
\end{eqnarray}
while ingoing solutions, i.e. those propagating towards the black hole, 
are characterized by
\begin{eqnarray}
\lim_{r^* \to +\infty}{|\Phi|} & \sim & 1 / r^4 \; ,\\
\lim_{r^* \to -\infty}{|\Phi|} & = & 0 \; . 
\end{eqnarray}

In the following figures we display the results obtained from runs of the
Teukolsky code described in Ref.\ \cite{KLPA97}
with grid spacings of $2 \,\delta t = \delta r^*  
=0.05,0.1$, and $\delta\theta=\pi/64,\pi/32$. 
(We have set the mass of the Schwarzschild
black hole to unity). 
For the comparisons with the Teukolsky evolution,
the ($1+1$) Zerilli equation (Eq.\ (\ref{rtzerilli}) with ${\cal S}_\ell
\equiv 0$) was evolved with the same temporal
and radial grid spacings as the Teukolsky equation.
Previously, convergence tests 
have been performed by looking at the
numerical errors for different steps of integration compared to exact solutions
obtained by plugging a convenient analytic function into the Teukolsky equation
and taking into account the resulting (artificial) source. In this way it is
shown that the convergence is (almost) quadratic.
 The code can be used for any value of the multipole
$\ell$, but in practice we used it only for $\ell=2$ and 4. This is so, because
for the head-on collision of equal masses holes only even values of $\ell$
appear, and the mode $\ell=2$ already contributes with over 90\% of the total
radiated energy. If one considers black holes of unequal masses, one would
obtain contributions from odd $\ell$, but still the radiation is strongly
dominated by the $\ell=2$ mode.
It should be noted that this consideration of pure multipoles in the
sense of the spherical harmonics is only possible for
nonrotating black holes: In the case of a rotating background, 
it is first of all impossible to generate initial data that represent
a pure multipole specified by $l$ and $m$. But even if we started the
evolution with a pure multipole, then other modes would be generated
in the course of the evolution. 
Therefore different multipoles will be present in the evolution.

Figure 1 depicts the waveforms for $\ell=2$ at the initial time $t=0$. We 
show together the initial Moncrief waveform $0.001 \times
\phi_{\text{M}} (t=0,r^*)$ for the evolution of the Zerilli equation
and the initial data for the Teukolsky equation $\Phi (t=0,r^*,\pi/2)$
and $\partial_t\Phi (t=0,r^*,\pi/2)$. (The observer is located in the
equatorial plane, i.e. $\theta=\pi/2$.)
Here it is worth to remark the fact that $\partial_t\Phi 
(t=0,r^*,\pi/2)\not=0$ even for time symmetric data (where 
$\partial_t\phi_{\text{M}}(t=0,r^*)=0)$, i.e. see Eq.\ 
(\ref{iniPhidotexplicit}).

In Figure 2 we depict the initial configuration 
for the $\ell=4$ mode.

Figure 3 shows the early behavior, as a function of the radial coordinate
$r^*$, of $\Phi (t=3,r^*,\pi/2)$ as computed in two different ways
with $\delta r^* =0.05$, and $\delta\theta=\pi/64$:
i) From the evolution of $\phi_{\text{M}}(t=0,r^*)$
with the Zerilli equation and then
using relation (\ref{chandra}) to build up $\Psi_4$ at the given time, and
ii) by directly evolving $\Psi_4 (t=0,r^*,\pi/2)$ with the
Teukolsky equation using the code of Ref.\ \cite{KLPA97} with the
parameter $a$ set to zero.
There is an agreement between the two evolution methods for $\ell=2$
waveforms within 2\% of the maximum amplitude (expected numerical error).

Figures 4 and 5 show the same comparison of waveforms at a later time
$(t=10$ in units of $M=1)$, for $\ell=2$ ($\delta r^*  
=0.1$, $\delta\theta=\pi/32$) and $\ell=4$ ($\delta r^*  
=0.05$, $\delta\theta=\pi/64$), respectively, in order
to check the agreement of the two methods of
computation for different multipoles.
We observe again that the curves accord within 2\% of the amplitude.

Figure 6 shows that the agreement of the waveforms evolved with the Zerilli
and Teukolsky equations continues to hold until late times. (Here we display
curves for $t=50$ obtained from an evolution with $\delta r^*  
=0.1$ and $\delta\theta=\pi/32$). Numerical errors are limited
to 2\% of the maximum amplitude of the waves.

These small errors have been obtained with short integration times and
not so small steps in the coordinates. If we take them as representative of the
errors in an astrophysically realistic computation we confirm both the
reliability and efficiency of the code we used for integrating the
Teukolsky equation.

Finally, in order to have a complete description (in the context of the
Teukolsky approach) of the close limit approach for head on collisions,
we show in Figure 7 the time dependence of $\Phi$ for $\ell=2$ as
it would be seen by an observer located at $r^*=50$, $\theta=\pi/2$.

\section{Discussion}

In this paper we reviewed the question of the imposition of Cauchy data
for the Teukolsky equation.
In order to write  the Weyl scalar $\Psi_4$ entering
into this wave equation in terms of pure hypersurface quantities,
we related $\Psi$ to the Moncrief waveform $\phi_{\text{M}}$, 
that already possesses the property of depending on the three-metric 
and the extrinsic curvature only. This relation was established 
via a Chandrasekhar transformation extended to the time domain.
As the Chandrasekhar transformation never makes use of any gauge fixing,
the transformation between waveforms (already gauge invariant)
and the subsequent results are gauge invariant. We checked analytically
that these results are equivalent to those obtained in Ref.\ \cite{CL98}. 
Using this relation to impose initial data
we have been able to reproduce the results of 
the close limit approximation\cite{PP94}, for the first time, by integrating
the Teukolsky equation instead of the Zerilli one.
In the astrophysically interesting case
of the close limit, we have also been able to check numerically,
that Chandrasekhar transformations give the same relations as found
in Ref.\ \cite{CL98}.
Furthermore, since these relations hold for any
hypersurface, we could test their correctness at
any time by evolving the
Zerilli and Teukolsky equations independently. 
We thus gained reliability about the final expressions obtained
in both, this paper and in Ref.\ \cite{CL98}.
In turn, we also gained further confidence on the numerical code developed in
Ref.\ \cite{KLPA97}, that focused on the late
time behavior of the radiation. Here 
the code is tested in the regime where the initial data influence is still
of some importance as opposed to the previous studies of the gravitational
tails.

All the tests that we performed have been made for perturbations around a
Schwarzschild, i.e.\ nonrotating, black hole. This allowed us the comparison
with the Zerilli-Moncrief formalism. In the general case of nonnegligible
rotation of the background hole, we do not have this counterpart to compare
with and one has no other option than to evolve the Teukolsky equation.
However, as we stated in the introduction, to the present day, we do not know
explicitly what initial data to evolve nor how to impose them to build up
$\Psi$ and $\partial_t\Psi$. There is perhaps an intermediate step we can
study with the present techniques which is the case of slow rotating holes that
still allow to be studied as perturbations of the Schwarzschild metric
\cite{GNPP98}.

\begin{acknowledgments}
We thank Pablo Laguna, Hans-Peter Nollert, Philippos Papadopoulos, and
Jorge Pullin for helpful discussions.
W.K. was supported by the NSF grant PHY-95-07719 and by research
funds of the University of Utah.
C.O.L. is a member of the Carrera del Investigador Cient\'\i fico of CONICET, 
Argentina and thanks FUNDACI\'ON ANTORCHAS for partial financial support.
\end{acknowledgments}

\newpage

\begin{figure}
\epsfbox{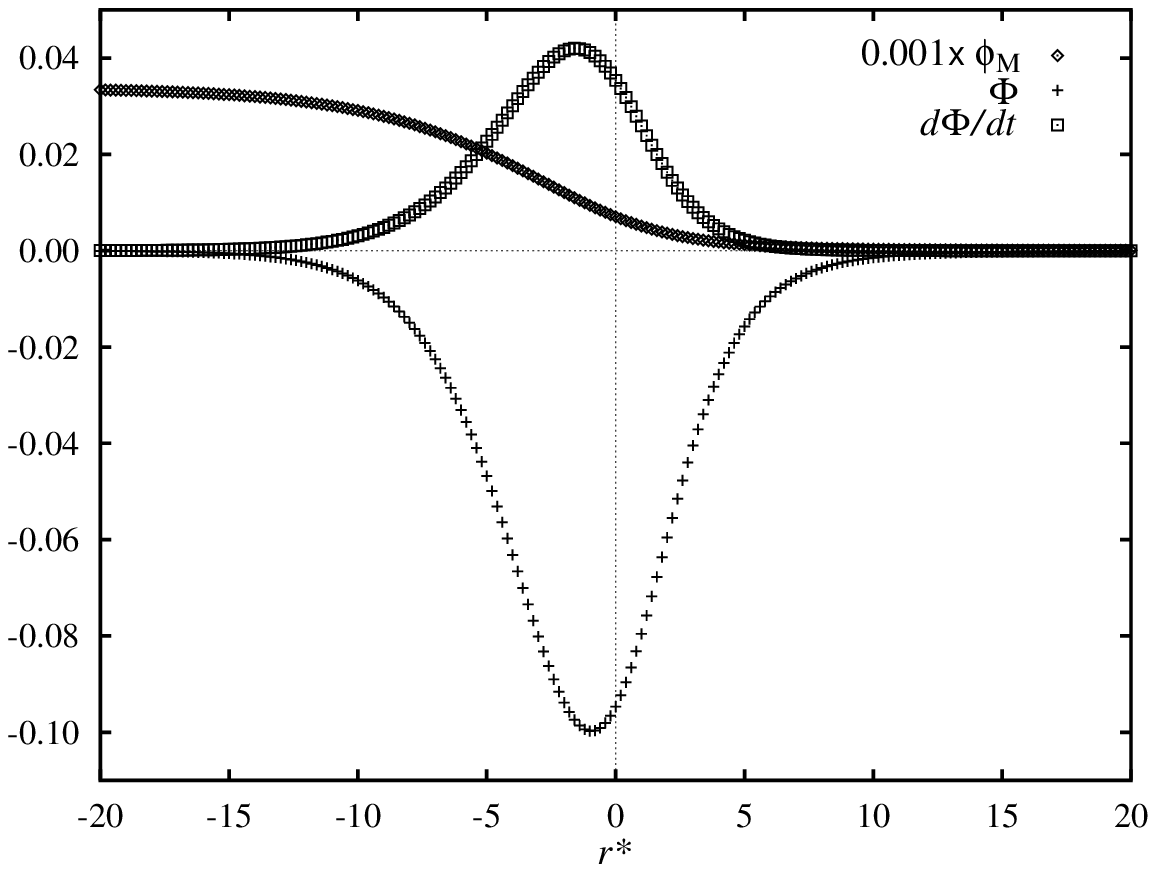}
\caption{
The radial dependence of the initial data for the Zerilli equation, 
$0.001 \times \phi_{\text{M}} (t=0,r^*)$, and for the Teukolsky equation, 
$\Phi(0,r^*,\pi/2)$, and $\partial_t\Phi(0,r^*,\pi/2)$, for the $\ell=2$ mode.}
\end{figure}
\begin{figure}
\epsfbox{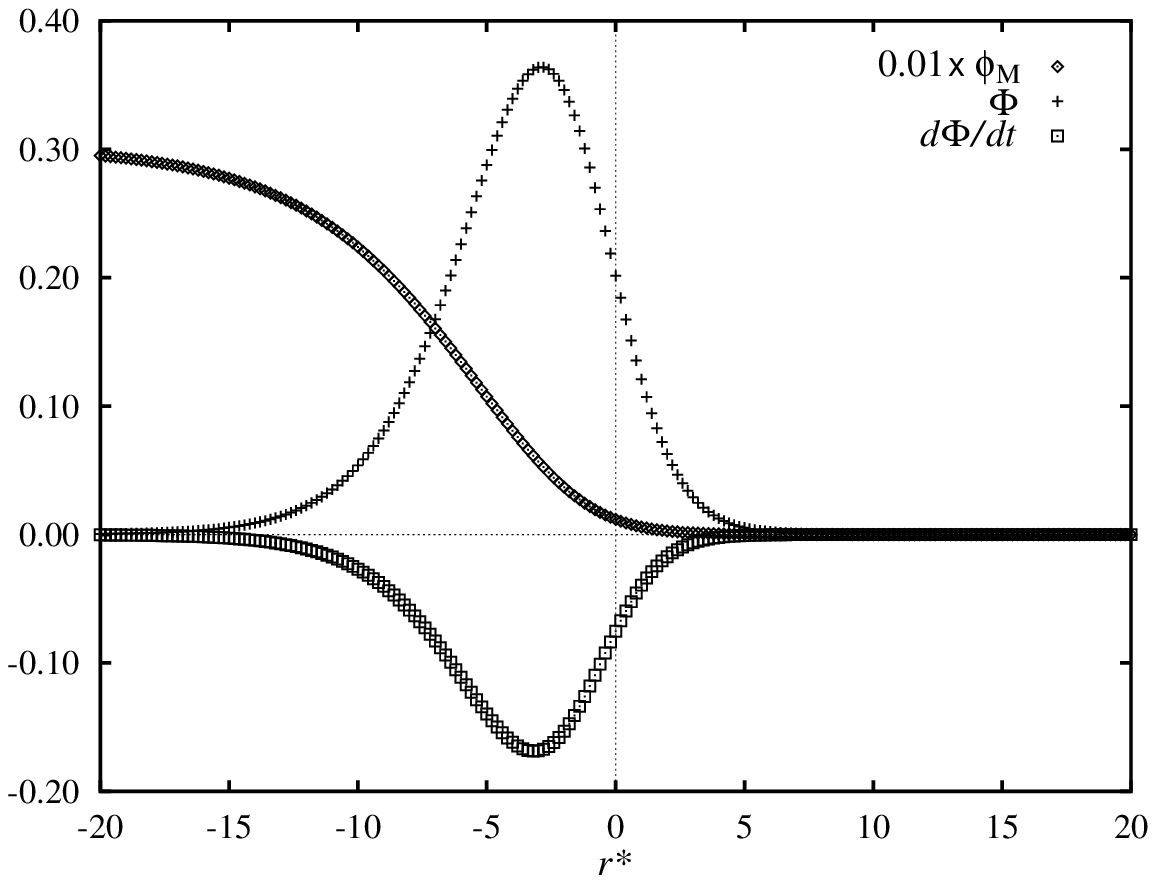}
\caption{
The radial dependence of the initial data for the Zerilli equation, 
$0.01 \times \phi_{\text{M}} (t=0,r^*)$, and for the Teukolsky equation, 
$\Phi(0,r^*,\pi/2)$, and $\partial_t\Phi(0,r^*,\pi/2)$, for the $\ell=4$ mode.}
\end{figure}
\begin{figure}
\epsfbox{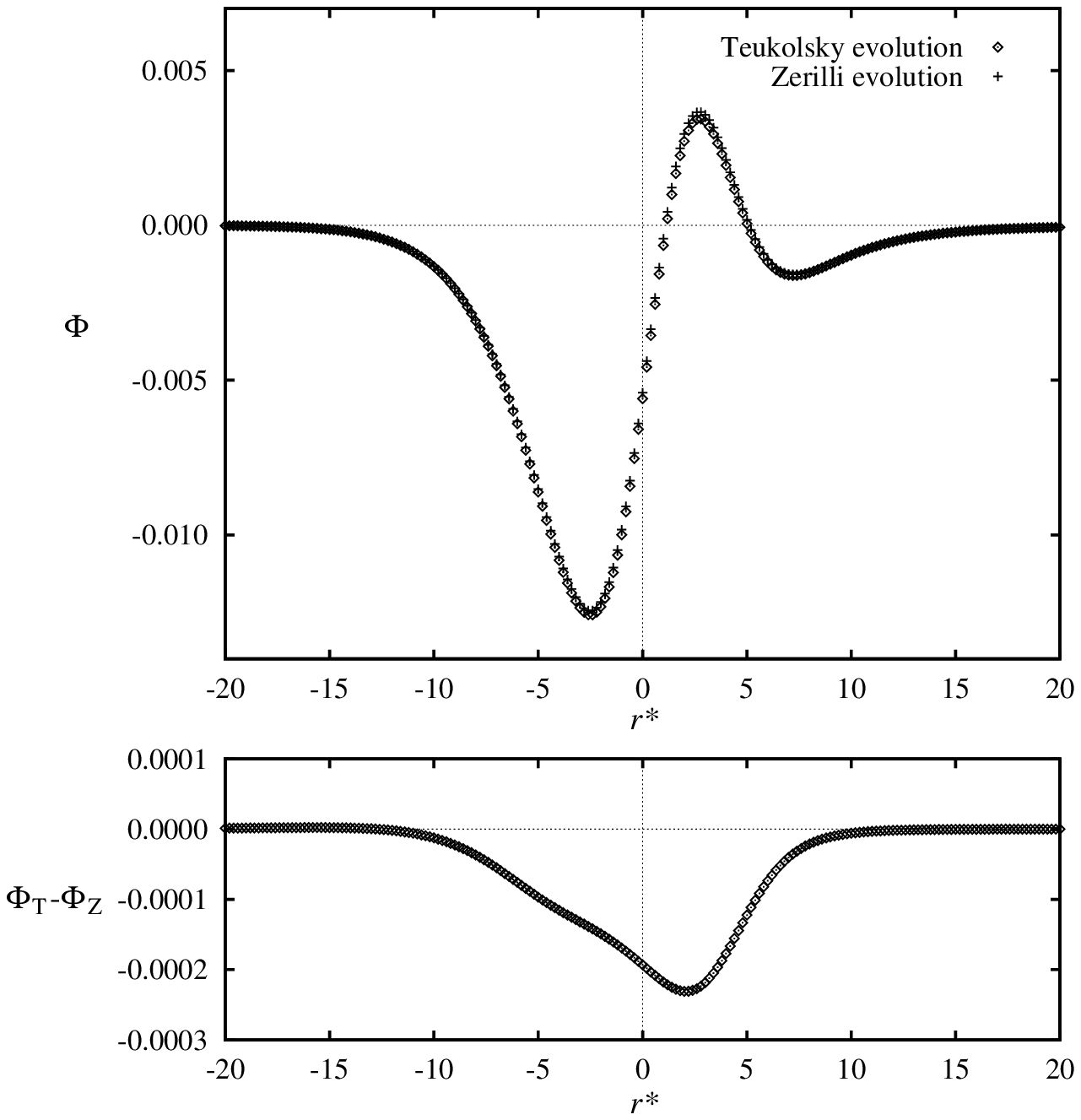}
\caption{Early radial dependence of $\Phi(t=3,r^*,\pi/2)$ for the $\ell=2$
multipole computed 
from the evolution of $\phi_{\text{M}} (t,r^*)$ via the numerical
integration of the Zerilli equation and the transformations
(\protect\ref{chandra}), and from the evolution of the Teukolsky code.}
\end{figure}
\begin{figure}
\epsfbox{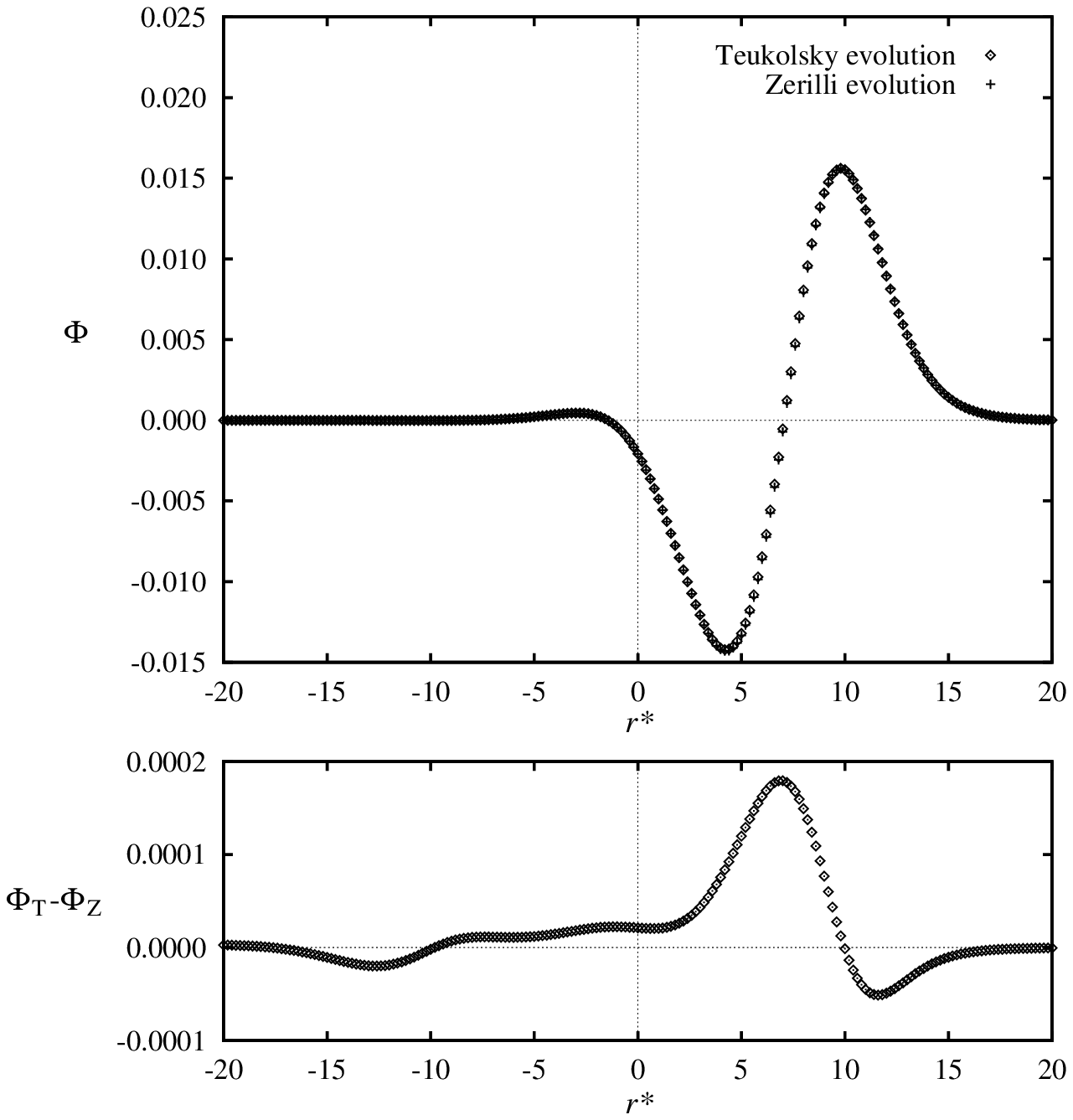}
\caption{Later radial behavior of $\Phi(t=10,r^*,\pi/2)$ for the $\ell=2$
multipole computed 
from the evolution of $\phi_{\text{M}} (t,r^*)$ via the numerical integration
of the Zerilli equation, and from evolving the Teukolsky code.}
\end{figure}
\begin{figure}
\epsfbox{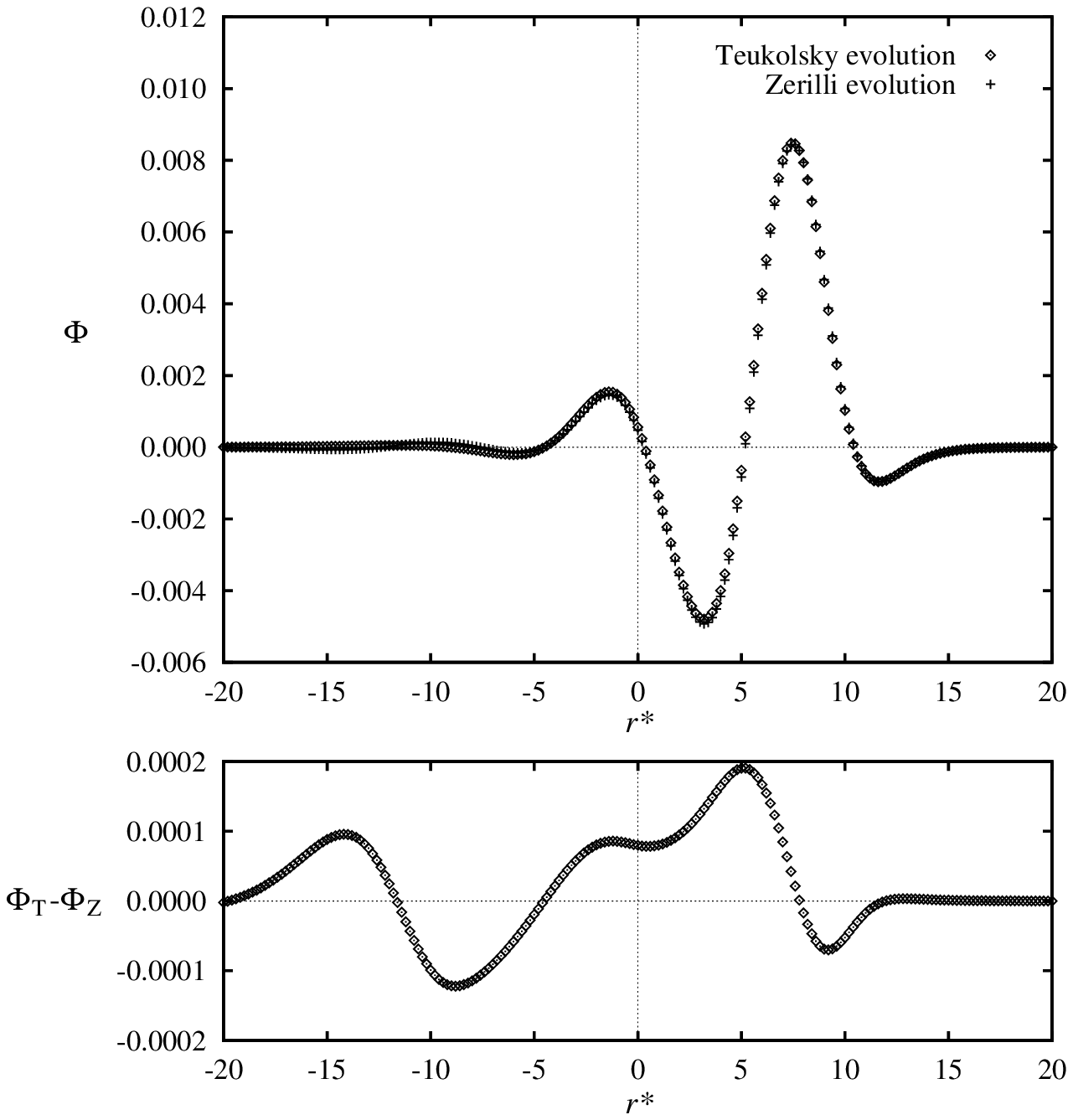}
\caption{Later radial behavior of $\Phi(t=10,r^*,\pi/2)$ for the $\ell=4$
multipole computed 
from the evolution of $\phi_{\text{M}} (t,r^*)$ via the numerical integration
of the Zerilli equation, and from the Teukolsky code evolution.}
\end{figure}

\begin{figure}
\epsfbox{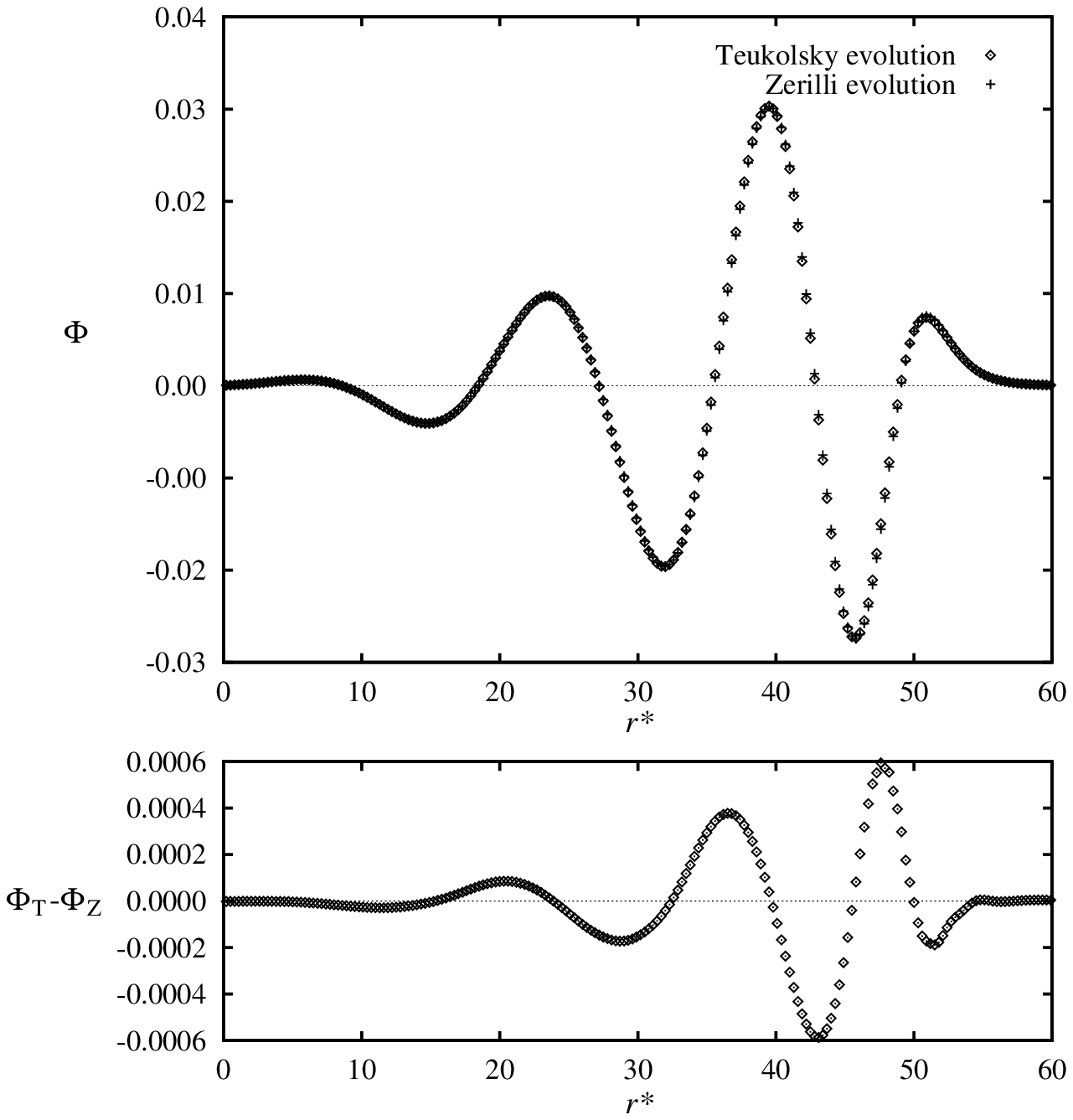}
\caption{Late radial behavior of $\Phi(t=50,r^*,\pi/2)$ for $\ell=2$
as computed through the integration of the Zerilli equation and through
the Teukolsky equation.}
\end{figure}

\begin{figure}
\epsfbox{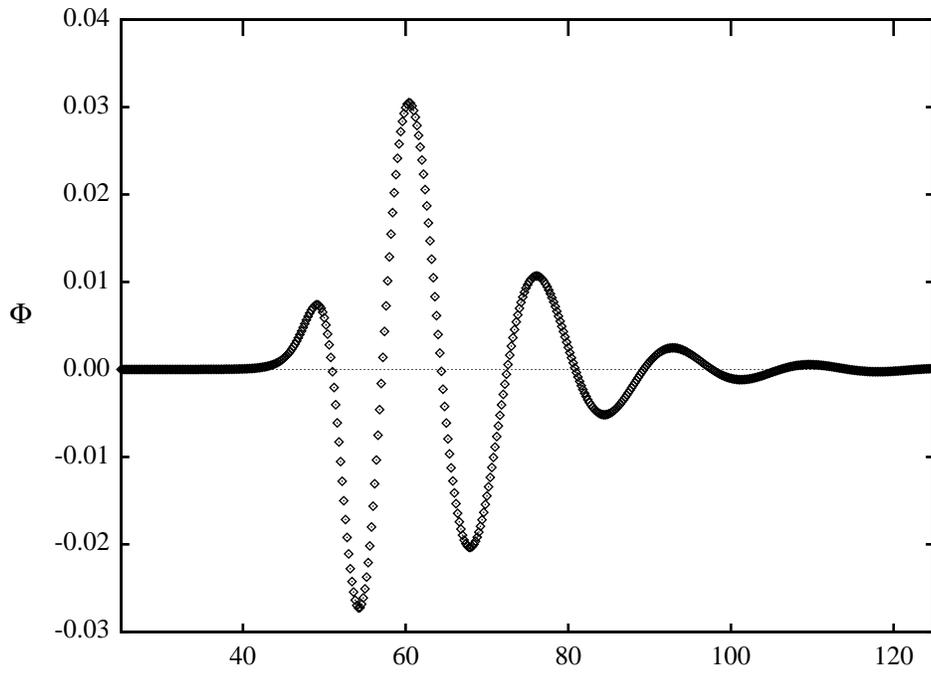}
\caption{Waveform for $\ell=2$. The function $\Phi$ is plotted as a
function of the coordinate time $t$, for a detector
located at $r^*=50$, $\theta=\pi/2$. 
}
\end{figure}

\end{document}